\documentclass[%
aps,
jmp,%
amsmath,amssymb,
reprint,%
]{revtex4-2}
\usepackage{graphicx}
\usepackage{subfigure}
\usepackage{dcolumn}
\usepackage{bm}
\usepackage{graphicx}

\usepackage{graphicx}
\usepackage{subfigure}
\usepackage{dcolumn}
\usepackage{bm}
\usepackage[mathlines]{lineno}
\graphicspath{{figure/}}
\usepackage{array}
\usepackage{braket}
\usepackage{diagbox}
\usepackage{amsmath}
\usepackage{appendix}
\usepackage{multirow}
\usepackage{bm,color,bbm}

\usepackage{float}
\newcolumntype{.}{D{.}{.}{-1}}
\usepackage{amsopn}
\usepackage[colorlinks,linkcolor=blue, anchorcolor=blue, urlcolor=blue, citecolor=blue]{hyperref}
\usepackage{epstopdf}

\begin{document}

	\title {Mode-pairing quantum key distribution with advantage distillation}
	\author{Xin Liu$^{1}$, Di Luo$^{1}$, Zhenrong Zhang$^{2}$, and Kejin Wei$^{1,*}$}
	\address{
		$^1$Guangxi Key Laboratory for Relativistic Astrophysics, School of Physical Science and Technology,
		Guangxi University, Nanning 530004, China\\		$^2$Guangxi Key Laboratory of Multimedia Communications and Network Technology, School of Computer Electronics and Information, Guangxi University, Nanning 530004, China\\
		$^*$Corresponding author: kjwei@gxu.edu.cn
	}

	\begin{abstract}
		
	Mode-pairing quantum key distribution (MP-QKD) is an easy-to-implement scheme that transcends the Pirandola--Laurenza--Ottaviani--Banchi bound without using quantum repeaters. In this paper, we present an improvement of the performance of MP-QKD using an advantage distillation method. The simulation results demonstrate that the proposed scheme extends the transmission distance significantly with a channel loss exceeding 7.6 dB. Moreover, the scheme tolerates a maximum quantum bit error rate of 8.9\%, which is nearly twice that of the original MP-QKD. In particular, as the system misalignment error increases, the expandable distance of the proposed scheme also increases. The proposed system is expected to promote the practical implementation of MP-QKD in a wide range of applications, particularly in scenarios involving high channel losses and system errors.

	\end{abstract}
	\maketitle
	\section{Introduction}
	With the development of quantum computers, classical encryption methods based on algorithm complexity~\cite{shor1994proceedings,grover1996fast} have demonstrated to be insufficient in guaranteeing communication security. Thus, quantum key distribution (QKD)~\cite{BB84}, which operates by distributing information-theoretic secure keys, has garnered significant attention in research. However, there is a significant gap between the ideal QKD model and realistic models, resulting in security vulnerabilities in realistic QKD systems~\cite{RevModPhys.92.025002}. 
	
	Improvement of the practical security of QKD systems has been researched for decades. To this end, one approach involves the construction of more realistic models to analyze the security of QKD systems~\cite{2004Gottesman,2014Tamaki,2018Qian,Sun_2021,PhysRevApplied.19.014048}. However, characterizing all devices in real-world systems is challenging. In contrast, device-independent QKD has been proposed to address all security loopholes induced by device imperfections~\cite{PhysRevLett.98.230501}. However, this protocol exhibits a low key rate and is difficult to implement because of the strict requirement on the detection efficiency of single-photon detectors. Fortunately, measurement-device-independent QKD (MDI-QKD)~\cite{Lo}, which closes all loopholes in detection devices, is relatively simple to implement with excellent performance. Significant experimental progress has been made with regard to this method~\cite{PhysRevApplied.14.011001,PhysRevX.10.031030,woodward2021gigahertz,PhysRevApplied.15.064016,Fan-Yuan:21,GU20222167}. 
	
	Currently, most MDI-QKD methods are called two-mode MDI-QKD because they encode single-sided key information in the relative phases of the coherent states of two orthogonal optical modes. A successful two-photon interference measurement is required to correlate the information encoded in the photons by Alice or Bob in a two-mode scheme. If the photon emitted by Alice or Bob is lost during transmission, coincidence detection cannot be performed for successful interference, rendering the restoration of the raw key information impossible. Thus, the requirement of coincidence detection of MDI-QKD is a critical factor limiting its transmission distance.
	
	In recent years, the performance of MDI-QKD has been improved from multiple perspectives, e.g., increasing the secret key rate and the transmission distance~\cite{PhysRevA.99.052325,PhysRevX.9.041012,PhysRevA.101.052318,PhysRevA.103.012402,Jiang:22}. Twin-field QKD (TF-QKD)~\cite{lucamarini2018overcoming}, which encodes information in a single optical mode, is the most representative improvement. In particular, TF-QKD transcends repeaterless secret key capacity bound~\cite{takeoka2014fundamental}, and more tightly, the Pirandola-Laurenza-Ottaviani-Banchi (PLOB) bound~\cite{pirandola2017fundamental}. Subsequently, variants of the TF-QKD protocol have been proposed, such as phase-matching QKD, in which key information is encoded in phase with the coherent states~\cite{PhysRevX.8.031043}; sending-or-not-sending TF-QKD, in which information is encoded in intensity~\cite{Wang2018,yu2019sending}; etc~\cite{PhysRevApplied.11.034053,Chistiakov:19,Grasselli_2019,PhysRevApplied.14.064070,Wang_2020,Li:21,zhou2023twin}. However, TF-QKD requires locking the frequency and phase of the coherent state and stabilizing the global phase, which inevitably complicates the implementation setup with peripheral hardware~\cite{minder2019experimental,PhysRevLett.123.100505,PhysRevLett.123.100506,fang2020implementation,chen2021twin,wang2022twin,clivati2022coherent,liu2023experimental}. This makes experimentation challenging and hinders the use of single-mode schemes in real-life applications. 
	
	Recently, a new variant of MDI-type QKD, called mode-pairing QKD (MP-QKD), has been proposed~\cite{zeng2022mode}. Similar to TF-QKD, MP-QKD transcends the PLOB bound, but it does not require the use of phase-locking technology. This feature enables MP-QKD to be implemented using a simpler setup than TF-QKD. More recently, the tight finite-key effect~\cite{wang2023tight} and experimental demonstrations using off-the-shelf optical devices for MP-QKD have been reported~\cite{PhysRevLett.130.030801}.
	
	In this study, we further improved the performance of MP-QKD by using the advantage distillation (AD) method. The proposed scheme modifies the post-processing step without changing the hardware of a realistic MP-QKD system. Hence, it can be directly applied to current systems. Its fundamental underlying concept is to divide the original key into blocks of a few bits each, enabling highly correlated keys to be distinguished from weakly correlated bits. The typical experimental parameters of MP-QKD are used for simulations. The simulation results demonstrate that the proposed scheme extends the transmission distance significantly. Moreover, the maximum tolerable quantum bit error rate (QBER) of the proposed system is 8.9\%, which is nearly twice that of the original MP-QKD. In particular, in some specific cases, the proposed scheme exhibits a longer expandable distance, paving the way for the widespread real-world application of MP-QKD.
	
	The remainder of this paper is organized as follows. In Sec.~\ref{Mode-pairing QKD}, we briefly summarize the steps involved in MP-QKD. In Sec.~\ref{Mode-pairing QKD with AD}, we introduce the protocol steps of the proposed scheme and present its security analysis in detail. Subsequently, in Sec.~\ref{SIMULATION}, we present the numerical simulation results, comparing the performances of the proposed scheme and the original MP-QKD. The results demonstrate the impact of the misalignment error $e_{d}$ on the performance of the proposed scheme. Finally, we present further discussion and our conclusions in Sec.~\ref{conclusion and discussion}.
	
	\section{Original MP-QKD}
	\label{Mode-pairing QKD}
	
\begin{figure}[h!]
		\centering
		\includegraphics[width=1\linewidth]{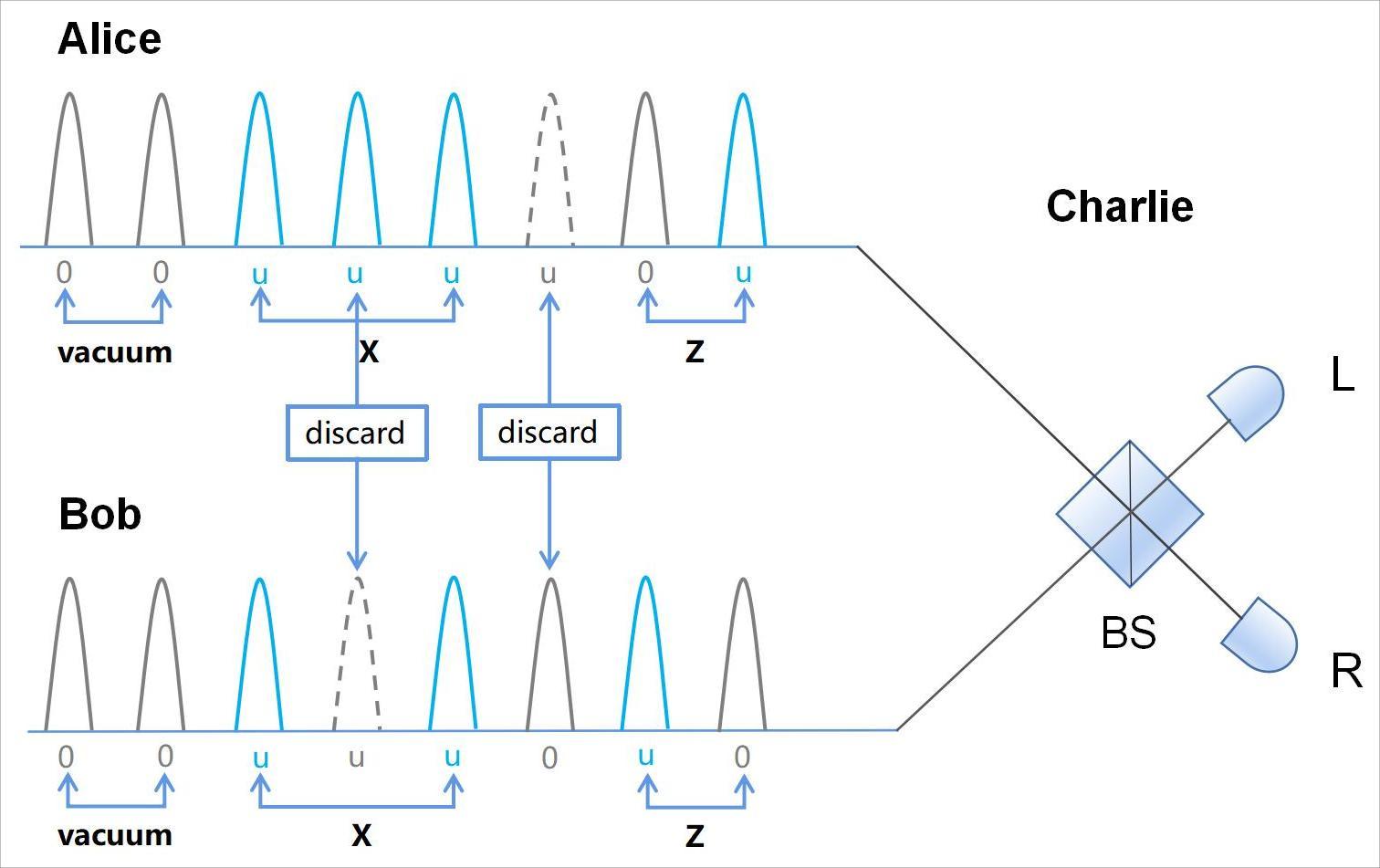}
		\caption{Diagram of the MP-QKD scheme. Alice and Bob transmit the prepared coherent pulses to Charlie, and they utilize the $Z$-pairs to generate the key after post-processing. For time bins $k$ and $l$, one intensity is 0 and the other is nonzero as $Z$-basis, the intensities are given by $\left ( \mu ,\mu  \right ) $ as the $X$-basis, and they are $(0, 0)$ in the vacuum state. If one or both of Alice and Bob lose data at the same time bin, they both discard the corresponding data and start searching for the following time bin. $L/R$: detector; BS: beam splitter.}
		\label{Mode-pairing-QKD}
	\end{figure}

	In this section, we briefly review the original MP-QKD method proposed in~\cite{zeng2022mode}. A schematic of this scheme is presented in Fig.~\ref{Mode-pairing-QKD} and its specific steps are summarized as follows:
	
	\itshape Step 1. Preparation. \upshape Alice (Bob) prepares $n$ weak coherent state pulses $\left|e^{\mathbf{i}\theta_{A}^{k}} \sqrt{\lambda _{A}^{k}}\right\rangle \left ( \left|e^{\mathbf{i}\theta_{B}^{k}} \sqrt{\lambda _{B}^{k}}\right\rangle\right ) $ with intensities $\lambda _{A}^{k} \left ( \lambda _{B}^{k} \right ) \in \left \{ \mu , 0 \right \} $, where each time bin satisfies $k\in \left \{ 1,2,\cdots ,n \right \} $, and the phase satisfies $\theta_{A}^{k} \left ( \theta_{B}^{k} \right ) \in[0,2 \pi)$.
	
	\itshape Step 2. Measurement and announcement. \upshape Alice and Bob transmit weak coherent light pulses to Charlie. For each time bin $k$, Charlie performs an interference measurement on the two received pulses and records the responses of detectors \itshape L \upshape and/or \itshape R\upshape. Subsequently, Charlie publicly announces whether a detection event has been acquired and the detector that has clicked.
	
	\itshape Step 3. Mode pairing. \upshape Alice and Bob repeat the two aforementioned steps $N$ times. Corresponding to each round with successful detection, only one detector click (\itshape L \upshape or \itshape R \upshape) round is retained. Alice and Bob group the two clicked rounds into pairs to determine the bases. The phases and intensities encoded in these two rounds form a data pair. The paired bases are then compared---they are retained if they satisfy the sifting conditions; otherwise, they are discarded.
	
	\itshape Step 4. Basis sifting. \upshape For time bins $k$ and $l$, if one of the intensities is 0 and the other is nonzero, the data are retained and recorded as $Z$-basis, if the intensities are $\left ( \mu ,\mu  \right ) $ in terms of $X$-basis, or if the intensities are $(0, 0)$ as vacuum state, then the rest of the data are discarded. Subsequently, Alice and Bob announce the bases and the sum of the intensities corresponding to the time bins $k$ and $l$. If the announced bases are identical and no 'discard’ is present, the bases are recorded, and the data are retained.
	
	\itshape Step 5. Key mapping. \upshape For each $Z$-pair at time bins $k$ and $l$, Alice sets her key to $\kappa_{A}=0$ if the intensity pair is $\left ( \lambda _{A}^{k} ,\lambda _{A}^{l} \right ) =\left ( \mu,0 \right )$. Alternatively, Alice sets her key to $\kappa_{A}=1$ if the intensity pair is $\left ( \lambda _{A}^{k} ,\lambda _{A}^{l} \right ) =\left ( 0,\mu  \right )$. For each $X$-pair at time bins $k$ and $l$, the key is extracted from the relative phase $\left ( \theta _{A}^{l} -\theta _{A}^{k} \right ) =\phi _{A} +\pi \kappa _{A} $, where the raw key bit is given by $\kappa _{A} =\left [ \left ( \left ( \theta _{A}^{l} -\theta _{A}^{k} \right )/\pi  \right ) \text{mod} \, 2  \right ]$, and the alignment angle is given by $\phi_{A} :=\left ( \theta _{A}^{l} -\theta _{A}^{k} \right ) \text{mod} \, \pi$. Similarly, Bob assigns a raw key bit $\kappa_{B}$ and determines $\phi_{B}$. For each $X$-pair, Alice and Bob announce the alignment angles, $\phi _{A}$ and $\phi _{B}$. If $\phi _{A} =\phi _{B}$, the data pairs are retained; otherwise, they are discarded.
	
	\itshape Step 6. Parameter estimation. \upshape Alice and Bob use the $Z$-pairs to generate a key. All the raw data obtained can be used to estimate the bit error rate $E_{(\mu, \mu)}^{ZZ}$ of the raw key in $Z$-pairs with overall intensities of $\left ( \lambda _{A}^{k,l} ,\lambda _{B}^{k,l} \right ) =\left ( \mu ,\mu  \right )$. When Alice and Bob both transmit a single photon each at time bins $k$ and $l$, they can estimate the fraction of clicked signals, $\bar{q}_{11}$, using the data of $Z$-pairs with different intensities and can estimate the single-photon phase error rate, $e_{\left ( 1,1 \right ) }^{XX} $, using the data of the $X$-pairs.
	
	\itshape Step 7. Post-processing. \upshape Alice and Bob perform error correction and privacy amplification on the raw key
	data to obtain the final secret key.
	
	Based on the security proof presented in~\cite{zeng2022mode}, the final key rate of MP-QKD can be estimated as follows:
	\begin{equation}
		R= r_{p}(p, \delta) r_{s}\left\{\bar{q}_{11}\left[1-H\left(e_{(1,1)}^{XX}\right)\right]-f H\left(E_{(\mu , \mu)}^{ZZ}  \right)\right\},\label{Rate}
	\end{equation}
	where $r_{p}(p,\delta)$ denotes the expected pair rate contributed during each round, $\delta$ denotes the maximum pairing interval, and $p$ denotes the probability of the $k$-th emitted pulse results in each successful click. $r_{s}$ denotes the probability that a generated pair is a $Z$-pair, $e_{(1,1)}^{XX}$ denotes the single-photon phase error rate, $\bar{q}_{11}$ denotes the expected single-photon pair ratio in all $Z$-pairs, $f$ denotes the error correction efficiency, and $E_{(\mu, \mu)}^{ZZ}$ denotes the bit error rate of the $Z$-pairs. The detailed calculation process for obtaining these parameters is described in Appendix~\ref{Appendix A}.
	
	\section{MP-QKD with AD method}
	\label{Mode-pairing QKD with AD}
	
	Next, we discuss the specific steps involved in applying the AD method to MP-QKD. The AD method~\cite{PhysRevLett.124.020502,li2022improving,li2022,Wang_2022,Zhu:23,Jiang:23,hu2023practical} only changes the post-processing steps. Thus, steps $1-6$ of the proposed scheme are identical to those of the original MP-QKD. The only change is in the post-processing procedure of the original MP-QKD. The specific details are presented below.
	
	\itshape New step 7. \upshape Alice and Bob divide their raw key into $b$ blocks each, i.e., $ \{x_{1}, x_{2}, \ldots, x_{b}\}$ and $\{y_{1}, y_{2}, \ldots, y_{b}\}$. Alice randomly selects a bit $c\in  \{ 0,1 \}$ and transmits the messages $m=\{m_{1}, m_{2}, \ldots, m_{b}\}=\{x_{1}\oplus c, x_{2}\oplus c, \ldots, x_{b}\oplus c\} $ to Bob via an authenticated classical channel. Alice and Bob accept the block only if Bob announces that the result of $\{m_{1}\oplus y_{1} , m_{2}\oplus y_{2}, \ldots, m_{b}\oplus y_{b}\} $ is $ \{ 0,0,\dots ,0 \} $ or $\{ 1,1,\dots ,1  \} $. Then, they retain the first bits, $x_{1}$ and $y_{1}$, as raw keys. Finally, Alice and Bob perform error correction and privacy amplification on the raw key data to obtain the secret keys.
			
	\begin{figure}[t]
		\centering
		\includegraphics[height=6.8cm,width=7.8cm]{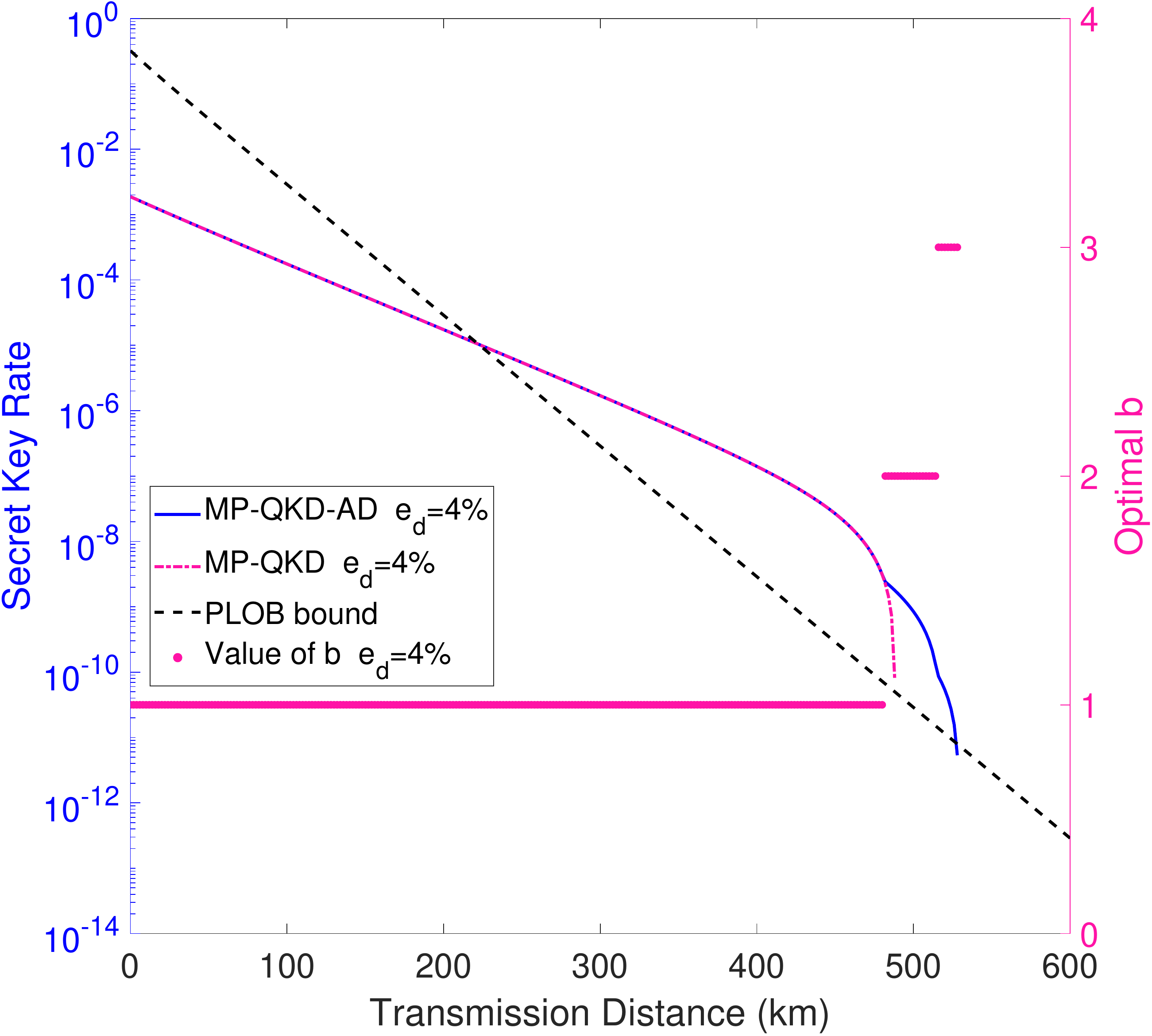}
		\caption{Performance comparison between the proposed scheme (MP-QKD-AD) and the original MP-QKD scheme and the relationship between the optimal $b$ values and the transmission distance assuming $e_{d}=4\%$. The blue solid line represents the secret key rate of the proposed scheme, the pink dotted line represents the secret key rate of the original MP-QKD scheme, the black dotted line represents PLOB bound, and the pink scattered points represent the optimized $b$ values of the proposed scheme.}
		\label{ed=0.04}
	\end{figure}
	
	To obtain further insights into the improvement of the achieved key rate using the AD method, we first reanalyze the key rate of MP-QKD using quantum information theory. We rewrite the key rate formulas as follows:
	
	\begin{equation}
		\begin{small}
		\begin{aligned}
			R \geq & \min _{\lambda_{0}, \lambda_{1}, \lambda_{2}, \lambda_{3}} r_{p}(p,\delta) r_{s}\left\{\overline { q } _ { 1 1 } \left[1-(\lambda_{0}+\lambda_{1}) h\left(\frac{\lambda_{0}}{\lambda_{0}+\lambda_{1}}\right)\right.\right. \\
			& \left.\left.-(\lambda_{2}+\lambda_{3}) h\left(\frac{\lambda_{2}}{\lambda_{2}+\lambda_{3}}\right)\right]-f h(E_{(\mu, \mu)}^{Z Z})\right\}
		\end{aligned}\label{key rate}
	\end{small}
	\end{equation}
	where $\sum_{j=0}^{3} \lambda_{j}=1$, and $\lambda_{j} \left ( j =\left \{0,1,2,3\right \} \right )$ denote factors of the characterizing quantum channel. The single-photon error rates in the $X$-basis and $Z$-basis are constrained by $\lambda_{1}+\lambda_{3}=e_{\left ( 1,1 \right ) }^{XX}$ and $\lambda_{2}+\lambda_{3}=e_{( 1,1 ) }^{ZZ}$, respectively. A detailed analysis of Eq.~(\ref{key rate}) is presented in Appendix~\ref{Appendix B}.
	
	\begin{figure}[t]
		\centering
		\includegraphics[height=6.8cm,width=7.3cm]{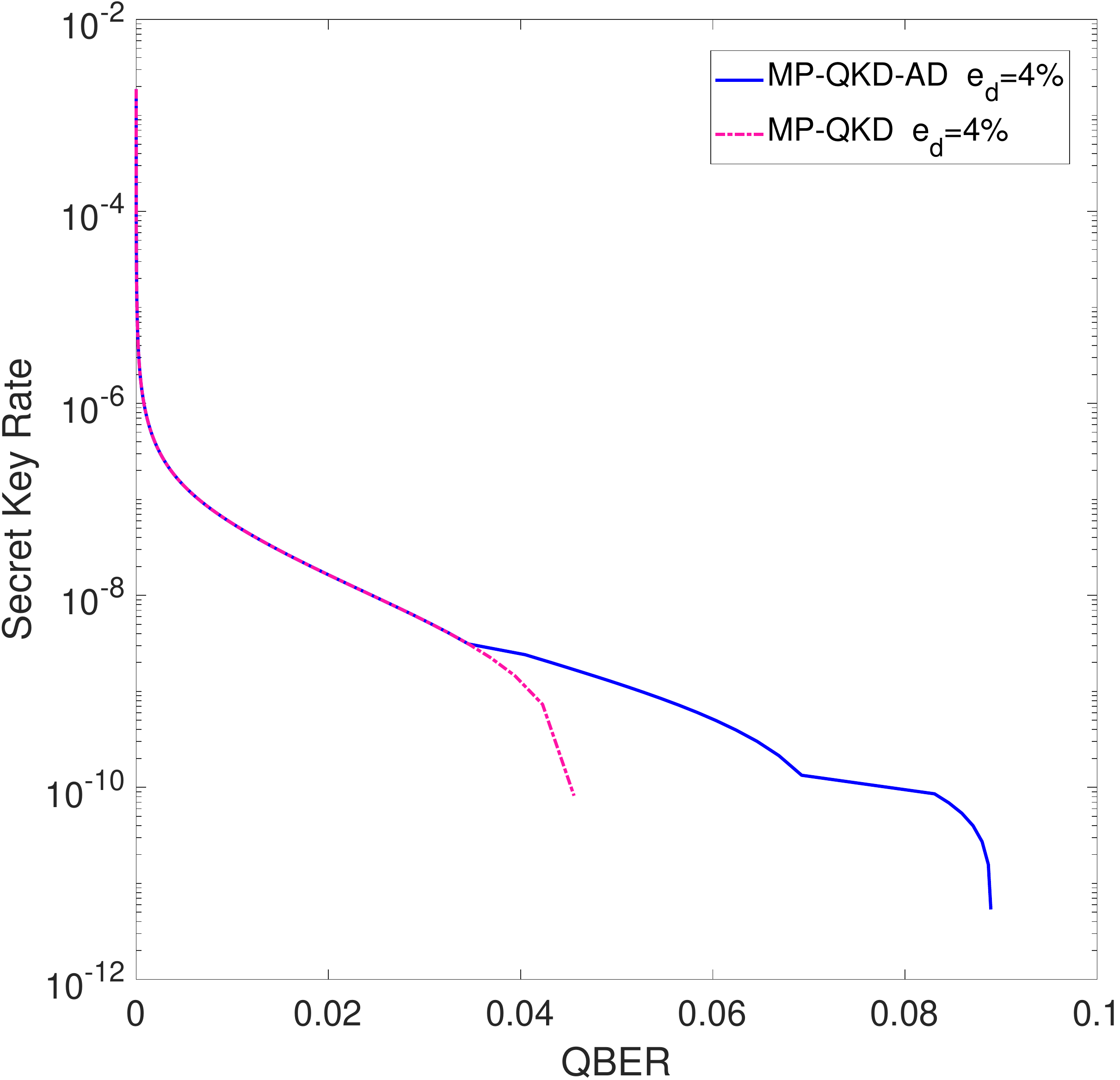}
		\caption{Comparison of the maximal tolerated quantum bit error rates (QBER) of the proposed scheme and the original MP-QKD assuming $e_{d}=4\%$. The blue solid line and pink dotted line represent the relationships between the secret key rate and the QBER of the proposed scheme and original MP-QKD, respectively.}
		\label{e_11}
	\end{figure}
	
	\begin{figure*}[t]
		\vspace{0.5cm} 
		\subfigbottomskip=2pt
		\subfigcapskip=1pt 
		\subfigure[]{
			\label{Different1}
			\includegraphics[height=210pt,width=230pt]{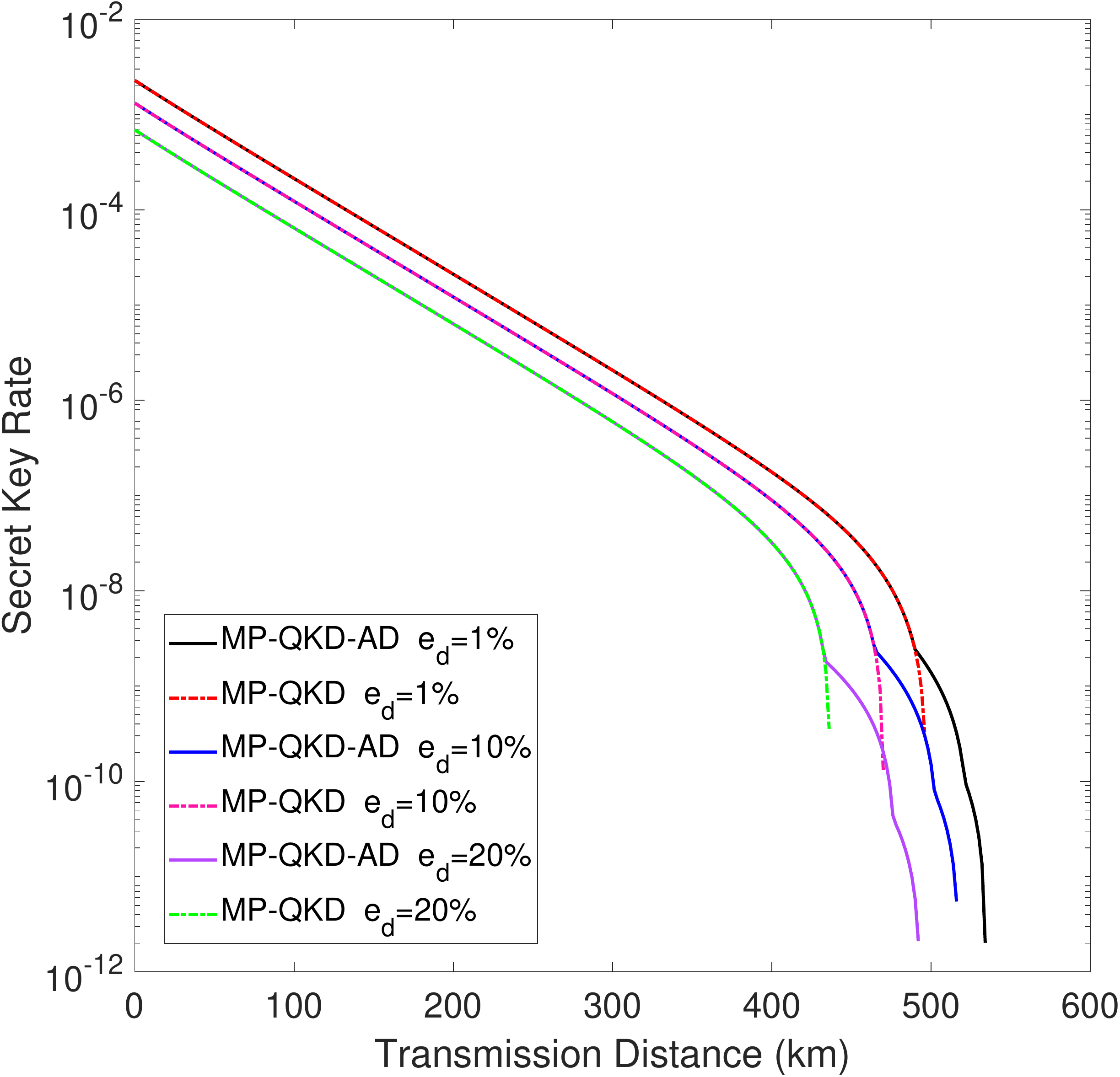}}
		\quad 
		\subfigure[]{
			\label{Different2}
			\includegraphics[height=210pt,width=225pt]{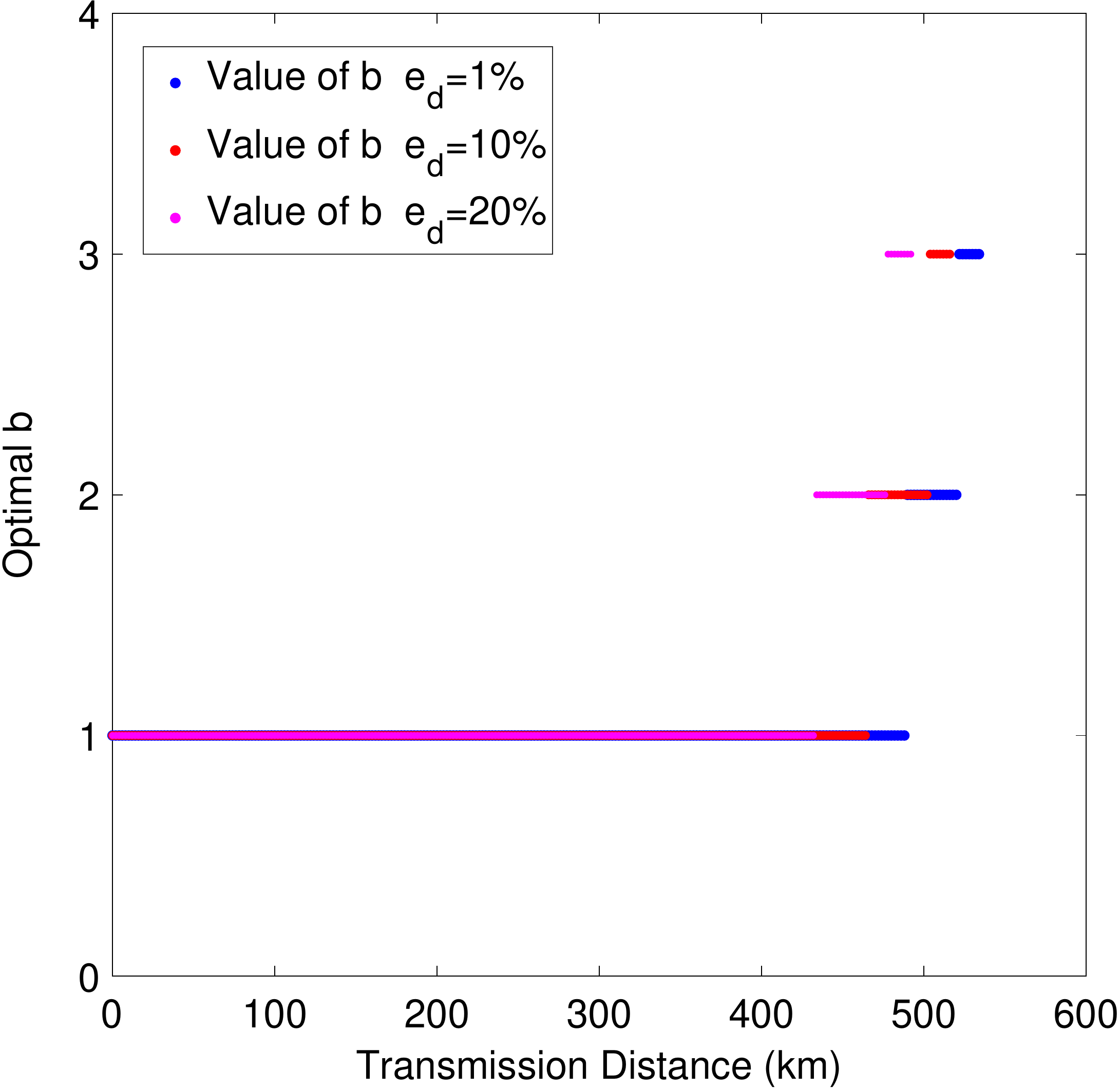}}
		
		\caption{The performance of the proposed scheme corresponding to different $e_{d}$. (a) Relationship between the secret key rate and the total transmission distance corresponding to different $e_{d}$. The solid lines and dotted lines of different colors represent the performances of the proposed scheme and the original MP-QKD corresponding to different $e_{d}$, respectively. (b) Relationship between the optimal value of b and the total transmission distance corresponding to different $e_{d}$. The blue, red, and pink scattered points represent the optimal value $b$ with respect to different distances corresponding to $e_{d}=1\%$, $10\%$, and $20\%$, respectively.}
		\label{Comparison}
	\end{figure*}
	
	After post-processing using the AD method (new step 7), highly correlated bits can be separated from weakly correlated information, and the key rate of the MP-QKD protocol can be modified as follows (the detailed formulas following AD post-processing are presented in Appendix~\ref{Appendix C}):  
	\begin{equation}
		\begin{aligned}
			\tilde{R} \geq & \max _{b} \min _{\lambda_{0}, \lambda_{1}, \lambda_{2}, \lambda_{3}} \frac{1}{b} q_{\text {s}} r_{p}(p,\delta) r_{s}\left\{(\bar{q}_{11})^{b}\right. \\
			& {\left[1-(\tilde{\lambda}_{0}+\tilde{\lambda}_{1}) h\left(\frac{\tilde{\lambda}_{0}}{\tilde{\lambda}_{0}+\tilde{\lambda}_{1}}\right)\right.} \\
			& \left.\left.-(\tilde{\lambda}_{2}+\tilde{\lambda}_{3}) h\left(\frac{\tilde{\lambda}_{2}}{\tilde{\lambda}_{2}+\tilde{\lambda}_{3}}\right)\right]-f h(\tilde{E}_{(\mu, \mu)}^{Z Z})\right\},
		\end{aligned} \label{R-AD}
	\end{equation}
	subject to
	\begin{equation}
		\begin{aligned}
			e_{\left ( 1,1 \right ) }^{XX}  &=\lambda_{1} +\lambda_{3},\\
			e_{\left ( 1,1 \right ) }^{ZZ} &=\lambda_{2} +\lambda_{3},\\
			q_{\text {s}} &=(E_{(\mu, \mu)}^{ZZ})^{b}+(1-E_{(\mu, \mu)}^{ZZ})^{b},\\
			\tilde{E}_{(\mu, \mu)}^{ZZ} & =\frac{(E_{(\mu, \mu)}^{ZZ})^{b}}{(E_{(\mu, \mu)}^{ZZ})^{b}+(1-E_{(\mu, \mu)}^{ZZ})^{b}},
		\end{aligned}
	\end{equation}
	and
	\begin{equation}
		\begin{aligned}
			\tilde{\lambda}_{0} & =\frac{\left(\lambda_{0}+\lambda_{1}\right)^{b}+\left(\lambda_{0}-\lambda_{1}\right)^{b}}{2 [\left(\lambda_{0}+\lambda_{1}\right)^{b}+\left(\lambda_{2}+\lambda_{3}\right)^{b}]}, \\
			\tilde{\lambda}_{1} & =\frac{\left(\lambda_{0}+\lambda_{1}\right)^{b}-\left(\lambda_{0}-\lambda_{1}\right)^{b}}{2 [\left(\lambda_{0}+\lambda_{1}\right)^{b}+\left(\lambda_{2}+\lambda_{3}\right)^{b}]}, \\
			\tilde{\lambda}_{2} & =\frac{\left(\lambda_{2}+\lambda_{3}\right)^{b}+\left(\lambda_{2}-\lambda_{3}\right)^{b}}{2 [\left(\lambda_{0}+\lambda_{1}\right)^{b}+\left(\lambda_{2}+\lambda_{3}\right)^{b}]}, \\
			\tilde{\lambda}_{3} & =\frac{\left(\lambda_{2}+\lambda_{3}\right)^{b}-\left(\lambda_{2}-\lambda_{3}\right)^{b}}{2 [\left(\lambda_{0}+\lambda_{1}\right)^{b}+\left(\lambda_{2}+\lambda_{3}\right)^{b}]},
		\end{aligned}
	\end{equation}
    where $q_{\text {s}}$ represents the probability of a successful advantage distillation on a block of length $b$, and $\tilde{E}_{(\mu, \mu)}^{ZZ}$ represents the overall error rate after the AD post-processing step.

	\section{Simulation}
	\label{SIMULATION}
    
	In this section, we report the simulation of the asymptotic performance of the proposed scheme using a typical symmetric quantum channel model and practical experimental parameters. The parameters for all numerical simulations described below are listed in Table~\ref{table1}. The parameters are adopted from Ref.~\cite{zeng2022mode}. The value of $b$ is restricted to the interval $\left [ 1,3 \right ]$.
	\begin{table}[h] \tiny
		\centering
		\caption{List of parameters used for numerical simulations. $\eta_{d}$ denotes the detection efficiency, $\alpha$ denotes the loss coefficient of the fiber, $p_{d}$ denotes detector dark count rate, $f$ denotes the error correction, and $\delta $ denotes the maximum pairing interval.}
		\resizebox{\linewidth}{!}{
			\begin{tabular}{cccccccccccc}
				\hline\hline
				$\eta_{d}$ &  $\alpha$ & $p_{d}$ & $f$ & $\delta $ \\ \hline
				20\%  &  0.2 dB/km & $1.2\times 10^{-8}$ & 1.15 & $10^{6}$ &   \\ 
				\hline\hline
			\end{tabular}
			\label{table1}
		}
	\end{table} 
	
	First, we specify the misalignment error $e_{d}$ for MP-QKD to be $4\%$ and compare the asymptotic secret key rate performance of the proposed scheme with that of the original MP-QKD. Fig.~\ref{ed=0.04} reveals that the performance of the proposed scheme is comparable to that of the original MP-QKD scheme corresponding to distances between 0 and 482 km. As the transmission distance increases, the secret key rate of the original MP-QKD decreases rapidly owing to the introduction of more noise, and the correlation of the original key deteriorates. When the distance exceeds 482 km, the proposed scheme represents an improvement over the original MP-QKD, and the maximum transmission distance increases by 40 km.
	
	To estimate the QBER tolerance of the proposed scheme, the relationship between the secret key rate and the QBER is simulated. The results are depicted in Fig.~\ref{e_11}. The simulation results indicate that QBER increases rapidly to 4.6\%, and the original MP-QKD becomes incapable of generating a secret key rate. In contrast, the proposed scheme remains capable of generating a secret key rate with a magnitude of $10^{-9} $. Thus, the proposed scheme tolerates a maximum QBER of 8.9\%, which is nearly twice that of the original MP-QKD.
		
	Finally, we investigate the effect of the misalignment error $e_{d}$ on the performance of the proposed scheme and the optimal $b$ values, with the results depicted in Fig.~\ref{Comparison}. When $e_{d}=1\%~\left ( e_{d}=10\%,~e_{d}=20\%\right ) $, the optimal $b$ value is greater than that at a distance of 490 km (466 km, 434 km), and the transmission distance of the proposed scheme increases by 38 km (46 km, 56 km). This observation implies that as the system misalignment error increases, the distance extension of the scheme also increases. 
	
	\section{Discussion and conclusions}
	\label{conclusion and discussion}
	In summary, this paper proposes a scheme to improve the performance of MP-QKD using the AD method and simulates its performance for an asymptotic case. The simulation results reveal that, compared to the original MP-QKD, the proposed scheme can tolerate a higher QBER and exhibits significantly increased transmission distance. When the QBER reaches 4.6\%, the original MP-QKD becomes incapable of generating the secret key rate. In contrast, the proposed scheme remains capable of generating a secret key rate with a magnitude of $10^{-9} $. The maximum QBER tolerated by the proposed scheme is nearly twice that of the original MP-QKD. Moreover, the expandable distance of high misalignment error systems in the proposed scheme is higher than that of low misalignment error systems. Thus, the proposed scheme outperforms the original MP-QKD in scenarios with high channel loss and system errors.
	
    The proposed scheme does not require any alterations to the original hardware devices---it only requires modification of the classical post-processing process. Thus, it can be applied to existing MP-QKD systems easily to improve their performance~\cite{PhysRevLett.130.030801}. 
		
    In future research, the possibility of improving the performance of MP-QKD further using random post-selection should be investigated as it has been shown to outperform the AD method in device-independent QKD~\cite{PhysRevLett.128.110506}. Moreover, the application of the AD method to asynchronous MDI-QKD protocol \cite{PRXQuantum.3.020315}, which is a similar single-mode MDI-type QKD, should be investigated.
	
	\section*{Acknowledgments}
	This study was supported by the National Natural Science Foundation of China (Nos. 62171144, 62031024, and  11865004), Guangxi Science Foundation (Nos.2021GXNSFAA220011 and 2021AC19384), and the Open Fund of IPOC (BUPT) (No. IPOC2021A02).

	\appendix 
	\section{SIMULATION MODEL OF MP-QKD }
	\label{Appendix A}
	
	According to Supplementary Note 4 in  Ref. ~\cite{zeng2022mode}, we can summarize the model of  mode-pairing QKD scheme as follows. 
	
	In the asymptotic case, we assume that the probability of Alice and Bob choosing a random emission intensity of $\left \{ 0,\mu   \right \} $ is close to $\frac{1}{2} $, and the probability of decoy intensity $\nu$ is negligible. We express the coherent pulse transmitted by Alice in the $k$-th round as $\left|\sqrt{\xi_{A}^{k} \mu}e^{\mathbf{i}\theta_{A}^{k}}\right\rangle$, where $\xi_{A}^{k}$ represents the random variable of intensity, $\theta_{A}^{k}$ is random phase. Similarly, Bob  transmits $\left|\sqrt{\xi_{B}^{k} \mu}e^{\mathbf{i}\theta_{B}^{k}}\right\rangle$ in the $k$-th round. The intensity setting for round $k$ is represented by the 2-bit vector $\xi^{k}:=\left[\xi_{A}^{k}, \xi_{B}^{k}\right]$.
	
	Alice and Bob are assumed to transmit weak coherent light pulses to Charlie through a typical symmetric-attenuation channel, the channel is i.i.d. for each round. Alice and Bob then pair the clicked pulses and determine their bases. For the $(k,l)$-th pulses to be paired, let $\tau^{k, l}=\left[\tau_{A}^{k, l}, \tau_{B}^{k, l}\right]:=\left[\xi_{A}^{k} \oplus \xi_{A}^{l}, \xi_{B}^{k} \oplus \xi_{B}^{l}\right]$, where $\oplus$ is the bit-wise addition modulo 2. When $\tau^{k, l}=\left[1,1\right]$ the $(k,l)$-pair is set to be a $Z$-pair.
	
	In the $k$-th round, we adopt two variables $\left ( L^{k},R^{k}\right )$ to represent the click events of the $L$ and $R$ detectors in the $k$-th round. The successful click variable is $ Cli^{k}=L^{k} \oplus R^{k}$. A successful click will occur only when $Cli^{k}=1 $. The detection probability $\operatorname{P}\left(Cli^{k}=1 \mid \xi^{k}\right)$ can be expressed as
	\begin{equation}
		\operatorname{P}\left(Cli^{k}=1 \mid \xi^{k}\right) \approx 1-\left(1-2 p_{d}\right) \exp \left[-\eta_{s} \mu\left(\xi_{A}^{k}+\xi_{B}^{k}\right)\right],
	\end{equation}
	
	The phase-randomized coherent states transmitted by $k$-th round can be regarded as a mixture of photon number states. $\operatorname{P}\left(Cli^{k}=1 \mid n^{k}\right)$ represents the detection probability when Alice and Bob respectively transmit photon number states $\left|n_{A}^{k}\right\rangle$ and $\left|n_{B}^{k}\right\rangle$, respectively, and is expressed as
	\begin{equation}
		\operatorname{P}\left(Cli^{k}=1 \mid n^{k}\right) \approx 1-\left(1-2 p_{d}\right)\left(1-\eta_{s}\right)^{\left(n_{A}^{k}+n_{B}^{k}\right)}.
	\end{equation}
	
	Next, we're going to consider the calculation of $r_{s}$. Without loss of generality, we regard the $k$-th and $l$-th rounds as a pair. For a general round, the probability of an intensity setting $\xi$ causing a click is given by
	\begin{equation}
		\operatorname{P}(\xi \mid Cli=1)=\frac{\operatorname{P}(\xi, Cli=1)}{\operatorname{P}(Cli=1)}=\frac{\operatorname{P}(Cli=1 \mid \xi)}{\sum_{\xi^{\prime}} \operatorname{P}\left(Cli=1 \mid \xi^{\prime}\right)}.
	\end{equation}
	Note that the subscripts are omitted because all rounds of detection are identical and independently distributed in our simulation.
	
	In the mode-pairing QKD scheme, a successful click occurs when $\tau ^{k,l} =\left [ 1,1 \right ] $. Therefore, four possible configurations of $\xi^{k}$ and $\xi^{l}$ (which generate $Z$-pairs) are 
	\begin{equation}
		\left[\xi^{k}, \xi^{l}\right] \in\{[00,11],[01,10],[10,01],[11,00]\},
	\end{equation}
	of these, $Err:=\{[00,11],[11,00]\}$ are the two configurations that cause bit error. To simplify the notation, we introduce several events,
	\begin{equation}
		\begin{array}{l}
			\operatorname{P}(Cli)=\operatorname{P}(\text { Pair Clicked }):=\operatorname{P}\left(Cli^{k}=Cli^{l}=1\right)=p^{2}, \\
			\operatorname{P}(E)=\operatorname{P}(\text { Pair Effective }):=\operatorname{P}\left(\xi^{k} \oplus \xi^{l}=11\right), \\
			\operatorname{P}(Err )=\operatorname{P}(\text { Pair Erroneous }):=\operatorname{P}\left(\left[\xi^{k}, \xi^{l}\right] \in Err \right), \\
			\operatorname{P}(S)=\operatorname{P}(\text { Single-photon Pair }):=\operatorname{P}\left(n^{k} \oplus n^{l}=11\right).
		\end{array}
	\end{equation}
	Below, we will list the possible situations of $\xi^{k}$ and $\xi^{l}$:
	\begin{equation}
		\begin{aligned}
			\xi^{k}&:=\left[\xi_{A}^{k}, \xi_{B}^{k}\right]=[0,1],[1,0],[0,0],[1,1] \\
			\xi^{l}&:=\left[\xi_{A}^{l}, \xi_{B}^{l}\right]=[0,1],[1,0],[0,0],[1,1]
		\end{aligned}
	\end{equation}
	of these, the conditions conforming to $\xi^{k} \oplus \xi^{l}=11$ are
	\begin{equation}
		\begin{aligned}
			\text { (1) } \xi^{k}:=\left[\xi_{A}^{k}, \xi_{B}^{k}\right]=[1,0] ~~ \xi^{l}:=\left[\xi_{A}^{l}, \xi_{B}^{l}\right]=[0,1] \\
			\text { (2) } \xi^{k}:=\left[\xi_{A}^{k}, \xi_{B}^{k}\right]=[0,1] ~~ \xi^{l}:=\left[\xi_{A}^{l}, \xi_{B}^{l}\right]=[1,0] \\
			\text { (3) } \xi^{k}:=\left[\xi_{A}^{k}, \xi_{B}^{k}\right]=[0,0] ~~ \xi^{l}:=\left[\xi_{A}^{l}, \xi_{B}^{l}\right]=[1,1] \\
			\text { (4) } \xi^{k}:=\left[\xi_{A}^{k}, \xi_{B}^{k}\right]=[1,1] ~~ \xi^{l}:=\left[\xi_{A}^{l}, \xi_{B}^{l}\right]=[0,0]
		\end{aligned}
	\end{equation}
	For case (1):
	\begin{equation}
		\begin{aligned}
			\operatorname{P}(Cli^{k}=1 \mid \xi^{k}) & \approx 1-(1-2 p_{d}) \exp \left[-\eta_{s} u (\xi_{A}^{k}+\xi_{B}^{k})\right] \\
			&=1-(1-2 p_{d}) \exp (-\eta_{s} u) \\
			\operatorname{P}\left(Cli^{l}=1 \mid \xi^{l}\right) & \approx 1-(1-2 p_{d}) \exp \left[-\eta_{s} u\left(\xi_{A}^{l}+\xi_{B}^{l}\right)\right]\\
			&=1-(1-2 p_{d}) \exp (-\eta_{s} u)
		\end{aligned}
	\end{equation}
	
	\begin{equation}
		\begin{aligned}
			\operatorname{P}(Cli^{k}=1 \mid n^{k}=\xi^{k}) &\approx 1-\left(1-2 p_{d}\right)\left(1-\eta_{s}\right)^{n_{A}^{k}+n_{B}^{k}} \\
			&=1-(1-2 p_{d})(1-\eta_{s}) \\
			\operatorname{P}\left(Cli^{l}=1 \mid n^{l}=\xi^{l}\right) &\approx 1-(1-2 p_{d})(1-\eta_{s})^{n_{A}^{l}+n_{B}^{l}} \\
			&= 1-(1-2 p_{d})(1-\eta_{s})
		\end{aligned}
	\end{equation}
	For case (2):
	\begin{equation}
		\begin{aligned}
			\operatorname{P}\left(Cli^{k}=1 \mid \xi^{k}\right) &\approx 1-(1-2 p_{d}) \exp (-\eta_{s} u) \\
			\operatorname{P}\left(Cli^{l}=1 \mid \xi^{l}\right) &\approx 1-(1-2 p_{d}) \exp (-\eta_{s} u)
		\end{aligned}
	\end{equation}
	
	\begin{equation}
		\begin{aligned}
			\operatorname{P}\left(Cli^{k}=1 \mid n^{k}=\xi^{k}\right) &\approx1-(1-2 p_{d})(1-\eta_{s}) \\
			\operatorname{P}\left(Cli^{l}=1 \mid n^{l}=\xi^{l}\right) &\approx1-(1-2 p_{d})(1-\eta_{s})
		\end{aligned}
	\end{equation}
	For case (3):
	\begin{equation}
		\begin{aligned}
			\operatorname{P}(Cli^{k}=1 \mid \xi^{k}) &\approx1-(1-2 p_{d})=2 p_{d} \\
			\operatorname{P}(Cli^{l}=1 \mid \xi^{l}) &\approx1-(1-2 p_{d}) \exp (-2 \eta_{s} u) \\
		\end{aligned}
	\end{equation}
	
	\begin{equation}
		\begin{aligned}
			\operatorname{P}(Cli^{k}=1 \mid n^{k}=\xi^{k}) &\approx 2 p_{d} \\
			\operatorname{P}(Cli^{l}=1 \mid n^{l}=\xi^{l}) &\approx 1-(1-2 p_{d})(1-\eta_{s})^{2}
		\end{aligned}
	\end{equation}
	For case (4):
	\begin{equation}
		\begin{aligned}
			\operatorname{P}(Cli^{k}=1 \mid \xi^{k}) &\approx 1-(1-2 p_{d}) \exp (-2 \eta_{s} u) \\
			\operatorname{P}(Cli^{l}=1 \mid \xi^{l}) &\approx 2 p_{d}
		\end{aligned}
	\end{equation}
	
	\begin{equation}
		\begin{aligned}
			\operatorname{P}(Cli^{k}=1 \mid n^{k}=\xi^{k}) &\approx 1-(1-2 p_{d})(1-\eta_{s})^{2} \\
			\operatorname{P}(Cli^{l}=1 \mid n^{l}=\xi^{l}) &\approx 2 p_{d}
		\end{aligned}
	\end{equation}
	
	The expected pair rate contributed during each round is
	\begin{equation}
		r_{p}(p,\delta)=\left[\frac{1}{p\left[1-(1-p)^{\delta}\right]}+\frac{1}{p}\right]^{-1},
	\end{equation}
	
	The expected successful click probability, i.e., the total transmittance of each round, is
	\begin{equation}
		\begin{small}
		\begin{aligned}
			&p:= \operatorname{P}\left(Cli^{k}=1\right)=\sum_{\xi^{k}} \operatorname{P} (Cli^{k}=1 \mid \xi^{k} ) \operatorname{P} (\xi^{k} ) \\
			&=\frac{1}{4} \sum_{\xi^{k}} \operatorname{P} (Cli^{k}=1 \mid \xi^{k} )\\
			&=\frac{1}{4} \left \{ 2 [ 1- ( 1-2p_{d} ) e^{-\eta _{s} \mu } ] +2p_{d} + [ 1-( 1-2p_{d} ) e^{-2\eta _{s} \mu } ]   \right \}\\ 
			&\approx \eta_{s}\mu.
		\end{aligned}
	\end{small}
	\end{equation}
	
	The $Z$-pair ratio $r_{s}$ is expressed as
	\begin{equation}
		\begin{small}
			\begin{aligned}
				&r_{s} = \operatorname{P}(E \mid Cli)=\operatorname{P}\left(\xi^{k} \oplus \xi^{l}=11 \mid Cli^{k}=1, Cli^{l}=1\right) \\
				&=\sum_{\xi^{k} \oplus \xi^{k}=11} \operatorname{P}\left(\xi^{k} \mid Cli^{k}=1\right) \operatorname{P}\left(\xi^{l} \mid Cli^{l}=1\right) \\
				&=\sum_{\xi^{k} \oplus \xi^{l}=11} \frac{\operatorname{P}\left(Cli^{k}=1 \mid \xi^{l}\right) \operatorname{P}\left(\xi^{k}\right)}{\operatorname{P}\left(Cli^{k}=1\right)} \frac{\operatorname{P}\left(Cli^{l}=1 \mid \xi^{l}\right) \operatorname{P}\left(\xi^{l}\right)}{\operatorname{P}\left(Cli^{l}=1\right)} \\
				&=\frac{1}{16} \frac{1}{p^{2}} \sum_{\xi^{k} \oplus \xi^{l}=11} \operatorname{P}\left(Cli^{k}=1 \mid \xi^{k}\right) \operatorname{P}\left(Cli^{l}=1 \mid \xi^{l}\right)\\
				&=\frac{1}{8} \frac{1}{p^{2}}  [ 1- ( 1-2p_{d} ) \exp ( -\eta _{s} \mu ) ]^{2}.
			\end{aligned}
		\end{small}
	\end{equation}
	
	The expected quantum bit error rate $E_{(\mu, \mu)}^{ZZ}$ of the $(k,l)$-pair is
	\begin{equation}
		\begin{aligned}
			E_{(\mu, \mu)}^{ZZ} &=\operatorname{P}(Err \mid E, Cli) \\
			&=\frac{\operatorname{P}(Err, E \mid Cli)}{\operatorname{Pr}(E \mid Cli)}=\frac{\operatorname{P}(Err \mid Cli)}{\operatorname{Pr}(E \mid Cli)} \\
			&=r_{s}^{-1} \operatorname{P}(Err \mid Cli).
		\end{aligned} \label{E}
	\end{equation}
	The erroneous pair event is included in the valid pair event. Therefore, the erroneous probability can be written as
	\begin{equation}
		\begin{footnotesize}
		\begin{aligned}
			&\operatorname{P}(E r r \mid Cli) =\operatorname{P}\left(\left[\xi^{k}, \xi^{l}\right] \in E r r \mid Cli^{k}=Cli^{l}=1\right) \\
			&=\sum_{\left[\xi^{k}, \xi^{l}\right] \in E r r} \operatorname{P}\left(\xi^{k} \mid Cli^{k}=1\right) \operatorname{P}\left(\xi^{l} \mid Cli^{l}=1\right) \\
			&=\sum_{\left[\xi^{k}, \xi^{l}\right] \in E r r} \frac{\operatorname{P}\left(Cli^{k}=1 \mid \xi^{k}\right) \operatorname{P}\left(\xi^{k}\right)}{\operatorname{P}\left(Cli^{k}=1\right)} \frac{\operatorname{P}\left(Cli^{l}=1 \mid \xi^{l}\right) \operatorname{P}\left(\xi^{l}\right)}{\operatorname{P}\left(Cli^{l}=1\right)} \\
			&=\frac{1}{16} \frac{1}{p^{2}} \sum_{\left[\xi^{k}, \xi^{l}\right] \in E r r} \operatorname{P}\left(Cli^{k}=1 \mid \xi^{k}\right) \operatorname{P}\left(Cli^{l}=1 \mid \xi^{l}\right).
		\end{aligned} \label{P_E}
	    \end{footnotesize}
	\end{equation}
	With the above Eq.~(\ref{E}) and Eq.~(\ref{P_E}), we can obtain
	\begin{equation}
		\begin{small}
		\begin{aligned}
			E_{(\mu, \mu)}^{ZZ}&=\frac{1}{16} \frac{1}{r_{s} p^{2}} \sum_{ [\xi^{k}, \xi^{l} ] \in E r r} \operatorname{P}(Cli^{k}=1 \mid \xi^{k}) \operatorname{P} (Cli^{l}=1 \mid \xi^{l} ) \\
			&=\frac{1}{4} \frac{1}{r_{s} p^{2}} p_{d} \left [ 1-\left ( 1-2p_{d}  \right ) \exp\left ( -2 \eta _{s} \mu  \right )  \right ].
		\end{aligned}
	\end{small}
	\end{equation}
	
	Then, we calculate the expected single-photon pair ratio $\bar{q}_{11}$ in the effective $Z$-pairs:
	\begin{equation}
		\begin{scriptsize}
			\begin{aligned}
				&\bar{q}_{11} =\operatorname{P}(S \mid E, Cli)=\frac{\operatorname{P}(S, E, Cli)}{\operatorname{P}(E, Cli)}=\frac{1}{r_{s} p^{2}} \operatorname{P}(S, E, Cli)\\ 
				&=\frac{1}{r_{s} p^{2}} \sum_{\xi^{k}, \xi^{l}} \operatorname{P}\left(S, E, Cli \mid \xi^{k}, \xi^{l}\right) \operatorname{P}\left(\xi^{k}, \xi^{l}\right) \\
				&=\frac{1}{16} \frac{1}{r_{s} p^{2}} \sum_{\xi^{k} \oplus \xi^{l}=11} \operatorname{P}\left(S, Cli \mid \xi^{k}, \xi^{l}\right) \\
				&=\frac{1}{16} \frac{1}{r_{s} p^{2}} \sum_{\xi^{k} \oplus \xi^{l}=11} \operatorname{P}\left(Cli \mid S, \xi^{k}, \xi^{l}\right) \operatorname{P}\left(S \mid \xi^{k}, \xi^{l}\right) \\
				&=\frac{1}{16} \frac{P_{\mu}(1)^{2}}{r_{s} p^{2}} \sum_{\xi^{k} \oplus \xi^{l}=11} \operatorname{P}\left(Cli^{k}=1 \mid n^{k}=\xi^{k}\right) \operatorname{P}\left(Cli^{l}=1 \mid n^{l}=\xi^{l}\right) \\
				&=\frac{1}{8} \frac{P_{\mu}(1)^{2}}{r_{s} p^{2}}\left [ 1-\left ( 1-2p_{d} \right ) \left ( 1-\eta _{s}  \right )   \right ] ^{2} .
			\end{aligned}
		\end{scriptsize}
	\end{equation}
	where $P_{\mu}(k)=\exp (-\mu) \frac{\mu^{k}}{k !}$ is the Poisson distribution, $\eta _{s} =\eta _{A} =\eta _{B} $, $\eta _{A}\left ( \eta _{B} \right )$ is the transmittance from Alice (Bob) to Charlie.	
	
	In the mode-pairing QKD scheme, if the decoy-state estimation is perfect, and the gain and error rate of  $X$-basis can be directly estimated by using the formula as follows~\cite{PhysRevA.86.062319}:
	\begin{equation}
			\begin{aligned}
				Y_{(1,1)} & =\left(1-p_{d}\right)^{2}\left[\frac{\eta_{A} \eta_{B}}{2}+\left(2 \eta_{A}+2 \eta_{B}-3 \eta_{A} \eta_{B}\right) p_{d}\right. \\
				& \left.+4\left(1-\eta_{A}\right)\left(1-\eta_{B}\right) p_{d}^{2}\right] \\
				e_{( 1,1 ) }^{XX} & =\frac{ [ e_{0} Y_{ ( 1,1  ) } -\left(e_{0}-e_{d}\right) (1-p_{d}^{2} ) \frac{\eta_{A} \eta_{B}}{2} ]}{Y_{ ( 1,1  ) }}
			\end{aligned}
	\end{equation}

	\section{SECURITY OF MP-QKD BASED ON QUANTUM INFORMATION THEORY}
	\label{Appendix B}
	
	The formula of key rate based on information theory is~\cite{renner2008security}
	\begin{equation}
		R=\min_{\sigma _{AB}\in \Gamma}S ( X\mid E ) -H ( X\mid Y ),
	\end{equation}
	where $\Gamma$ is the ensemble of all density operators $\sigma _{AB}$ on the $2\times 2 $-dimensional Hilbert space $\mathcal{H}_{A} \otimes \mathcal{H}_{B}$, $S\left ( X\mid E \right )$ is the von Neumann entropy, indicating the uncertainty of the eavesdropper's (Eve's) auxiliary state $E$ for the Alice's measurement result $X$, and $H\left ( X\mid Y \right )$ is the Shannon entropy, indicating the uncertainty of the receiver Bob's measurement result $Y$ to Alice's measurement result $X$.
	
	Similar to the security analysis based on the entanglement purification protocol, Alice and Bob prepare quantum states, $|1, 0\rangle^{k, l}$ and $|0, 1\rangle^{k, l}$are the eigenstates in the $Z$-basis, $| + \rangle=\left(|1,0\rangle^{k,l} + |0,1\rangle^{k, l}\right) / \sqrt{2}$ and $| - \rangle=\left(|1,0\rangle^{k,l} - |0,1\rangle^{k, l}\right) / \sqrt{2}$are the eigenstates in the $X$-basis, then sned them to Charlie for Bell-state measurement, where $\left |1,0\right \rangle ^{k,l} =\left | 1 \right \rangle ^{k} \left | 0 \right \rangle ^{l} $ indicates that there is one photon in time-bin $k$ and zero photons in time-bin $l$. Before Alice and Bob measure the quantum states, the whole system can be described by the following quantum state
	\begin{equation}
		|\Psi\rangle_{ABE} :=\sum_{j=0}^{3} \sqrt{\lambda_{j}} |\Phi  _{j} \rangle_{A B} \otimes |e_{j} \rangle_{E},
	\end{equation}
	where
	\begin{equation}
		\begin{aligned}
			| \Phi _{0}  \rangle =\frac{1}{\sqrt{2} }  ( |1, 0\rangle_{A}^{k,l} |1, 0\rangle_{B}^{k,l}+ |0, 1\rangle_{A}^{k,l} |0, 1\rangle_{B}^{k,l} ) ,\\
			| \Phi _{1}  \rangle =\frac{1}{\sqrt{2} }  ( |1, 0\rangle_{A}^{k,l} |1, 0\rangle_{B}^{k,l} - |0, 1\rangle_{A}^{k,l} |0, 1\rangle_{B}^{k,l} ) ,\\
			| \Phi _{2}  \rangle =\frac{1}{\sqrt{2} }  ( |1, 0\rangle_{A}^{k,l} |0, 1\rangle_{B}^{k,l}+ |0, 1\rangle_{A}^{k,l} |1, 0\rangle_{B}^{k,l} ),\\
			| \Phi _{3}  \rangle =\frac{1}{\sqrt{2} }  ( |1, 0\rangle_{A}^{k,l} |0, 1\rangle_{B}^{k,l} - |0, 1\rangle_{A}^{k,l} |1, 0\rangle_{B}^{k,l} ) ,
		\end{aligned}
	\end{equation}
	and $\sum_{j=0}^{3} \lambda_{j}=1$, $\lambda_{j} \left ( j =\left \{0,1,2,3\right \} \right )$ characterize quantum channel, the single-photon error rates in the $X$-basis and $Z$-basis are constrained by $\lambda_{1}+\lambda_{3}=e_{x}$ and $\lambda_{2}+\lambda_{3}=e_{z}$, respectively. The subscript $A$ denotes mode 'Alice' and $B$ denotes mode 'Bob'. $\left|e_{j}\right\rangle_{E}$ is an orthonormal basis of a 4-dimensional Hilbert space $\mathcal{H}_{E}$.
	
	Since the quantum channel is controlled by Eve, when measurement results of Alice and Bob are $00, 11, 01, 10$, respectively, the quantum states that Eve can obtain are
	\begin{equation}
		\begin{aligned}
			|\varphi_{0,0}\rangle=\frac{1}{\sqrt{2}}(\sqrt{\lambda_{0}}|e_{0}\rangle+\sqrt{\lambda_{1}}|e_{1}\rangle), \\
			|\varphi_{1,1}\rangle=\frac{1}{\sqrt{2}}(\sqrt{\lambda_{0}}|e_{0}\rangle-\sqrt{\lambda_{1}}|e_{1}\rangle), \\
			|\varphi_{0,1}\rangle=\frac{1}{\sqrt{2}}(\sqrt{\lambda_{2}}|e_{2}\rangle+\sqrt{\lambda_{3}}|e_{3}\rangle), \\
			|\varphi_{1,0}\rangle=\frac{1}{\sqrt{2}}(\sqrt{\lambda_{2}}|e_{2}\rangle-\sqrt{\lambda_{3}}|e_{3}\rangle),
		\end{aligned}
	\end{equation}
	Note that Eve can choose the optimal parameter $\lambda_{j} \left ( j =\left \{0,1,2,3\right \} \right )$ to reduce the security key rate, but $\lambda_{j}$ is constrained by the quantum bit error rate of two different bases.
	
	After the interference of Eve in the quantum channel, Alice and Bob obtain the density operators $\sigma_{X Y E}$ of the whole system by the orthonormal measurement of $\mathcal{H}_{A}$ and $\mathcal{H}_{B}$
	\begin{equation}
		\sigma_{X Y E}=\sum_{x, y}|x\rangle\langle x|\otimes| y\rangle\langle y| \otimes | \varphi_{x,y}\rangle\langle \varphi_{x,y}|
	\end{equation}
	
	Based on the above analysis, we can easily obtain
	\begin{equation}
		\begin{aligned}
			H(\sigma_{X E})=&1+h(\lambda_{0}+\lambda_{1}), \\
			H(\sigma_{E})=&h(\lambda_{0}+\lambda_{1})+(\lambda_{0}+\lambda_{1}) h(\frac{\lambda_{0}}{\lambda_{0}+\lambda_{1}})+\\
			&(\lambda_{2}+\lambda_{3}) h(\frac{\lambda_{2}}{\lambda_{2}+\lambda_{3}}), \\
			H(X \mid Y)=&h(\lambda_{0}+\lambda_{1}).
		\end{aligned} \label{H}
	\end{equation}
	
	Therefore, the final formula of key rate is
	\begin{equation}
		\begin{aligned}
			R \geqslant & \min _{\lambda_{0}, \lambda_{1}, \lambda_{2}, \lambda_{3}} S(X \mid E)- H(X \mid Y) \\
			= & \min _{\lambda_{0}, \lambda_{1}, \lambda_{2}, \lambda_{3}} H (\sigma_{X E} )-H (\sigma_{E} )-H (X \mid Y) \\
			= & \min _{\lambda_{0}, \lambda_{1}, \lambda_{2}, \lambda_{3}} 1- (\lambda_{0}+\lambda_{1}) h \left (\frac{\lambda_{0}}{\lambda_{0}+\lambda_{1}} \right ) \\
			& - (\lambda_{2}+\lambda_{3} ) h \left (\frac{\lambda_{2}}{\lambda_{2}+\lambda_{3}} \right )-h (\lambda_{0}+\lambda_{1} )
		\end{aligned}
	\end{equation}
	where $ h(x)=-x \log _{2}(x)-(1-x) \log _{2}(1-x)$ is the binary Shannon entropy function.
	
	To find the minimum value of the above key rate formula, the following conditions must be satisfied for $\lambda_{0}$, $\lambda_{1}$, $\lambda_{2}$ and $\lambda_{3}$ respectively
	\begin{equation}
		\begin{array}{l}
			\lambda_{0}=1-2 Q+\lambda_{3}, \\
			\lambda_{1}=Q-\lambda_{3}, \\
			\lambda_{2}=Q-\lambda_{3}, \\
			\lambda_{3}=Q^{2}.
		\end{array}
	\end{equation}
	where $Q$ is the bit error rate.
	
	In practical MP-QKD systems, phase-randomized weakly coherent sources are widely used to prepare quantum states. According to the specific protocol steps in Sec. ~\ref{Mode-pairing QKD}, Alice and Bob only use the successfully paired $Z$-basis to generate a key, and use the decoy state method to resist photo number splitting attacks. Among, all errors are corrected by Alice and Bob, so that in the Eq.~(\ref{H}) $H(X\mid Y) \leq f H (E_{ ( \mu ,\mu ) }^{ZZ} )$. $H(E_{( \mu ,\mu ) }^{ZZ} )$ is the maximum information that Eve steals during the error correction step. Therefore, the secret key rate of MP-QKD protocol can be given by
	\begin{equation}
		\begin{small}
		\begin{aligned}
			R \geq & \min _{\lambda_{0}, \lambda_{1}, \lambda_{2}, \lambda_{3}} r_{p}(p, \delta) r_{s}\left\{\bar { q } _ { 1 1 } \left[1-(\lambda_{0}+\lambda_{1}) h\left(\frac{\lambda_{0}}{\lambda_{0}+\lambda_{1}}\right)\right.\right. \\
			& \left.\left.-(\lambda_{2}+\lambda_{3}) h\left(\frac{\lambda_{2}}{\lambda_{2}+\lambda_{3}}\right)\right]-f h(E_{(\mu, \mu)}^{Z Z})\right\}
		\end{aligned} \label{R-information}
	\end{small}
	\end{equation}
	where $r_{p}(p,\delta)$ is the expected pair rate contributed during each round, $r_{s}$ is the probability that a generated pair is a $Z$-pair, and $\bar{q}_{11}$ is the expected single-photon pair ratio in all $Z$-pairs. $f$ is the error-correction efficiency, $E_{(\mu, \mu)}^{ZZ}$ is the bit-error rate of the $Z$-pairs.
	
	\section{SECURITY OF MP-QKD WITH AD}
	\label{Appendix C}
	
	In this section, we calculate the parameters in Eq.~(\ref{R-AD}) to estimate the secret key rate. In our protocol, Alice and Bob divide their raw key into blocks of $b$ size $\left\{x_{1}, x_{2}, \ldots, x_{b}\right\}$ and $\left\{y_{1}, y_{2}, \ldots, y_{b}\right\}$. Alice depends on a randomly chosen bit $c\in \left \{ 0,1 \right \}$ and sends the message $m=\left\{m_{1}, m_{2}, \ldots, m_{b}\right\}=\left\{x_{1}\oplus c, x_{2}\oplus c, \ldots, x_{b}\oplus c\right\} $ to Bob through an authenticated classical channel. They accept the block if and only if Bob announces the result of $\left\{m_{1}\oplus y_{1} , m_{2}\oplus y_{2}, \ldots, m_{b}\oplus y_{b}\right\} $ is either $\left \{ 0,0,\dots ,0 \right \} $ or $\left \{ 1,1,\dots ,1 \right \} $. By a straightforward calculation, the probability of a successful advantage distillation on a block of length $b$ is
	\begin{equation}
		q_{\text {s }} =\left(E_{(\mu, \mu)}^{ZZ}\right)^{b}+\left(1-E_{(\mu, \mu)}^{ZZ}\right)^{b}.
	\end{equation}

	For the message $m=\left\{m_{1}, m_{2}, \ldots, m_{b}\right\}$, once Eve gets any one of the measurements in $\left\{m_{1}, m_{2}, \ldots, m_{b}\right\}$, she (he) can get all $b$ measurements. Therefore, it can only be used to generate the secret key if all the $b$ pulses used for pairing are single-photon state, and the probability is $\left(\bar{q}_{11}\right)^{b}$. 
	
	With these, Eq.~(\ref{R-information}) can be modified as
	\begin{equation}
		\begin{aligned}
			\tilde{R} \geq & \max _{b} \min _{\lambda_{0}, \lambda_{1}, \lambda_{2}, \lambda_{3}} \frac{1}{b} q_{\text {s }} r_{p}(p, L) r_{s}\left\{(\bar{q}_{11})^{b}\right. \\
			& {\left[1-(\tilde{\lambda}_{0}+\tilde{\lambda}_{1}) h\left(\frac{\tilde{\lambda}_{0}}{\tilde{\lambda}_{0}+\tilde{\lambda}_{1}}\right)\right.} \\
			& \left.\left.-(\tilde{\lambda}_{2}+\tilde{\lambda}_{3}) h\left(\frac{\tilde{\lambda}_{2}}{\tilde{\lambda}_{2}+\tilde{\lambda}_{3}}\right)\right]-f h(\tilde{E}_{(\mu, \mu)}^{Z Z})\right\},
		\end{aligned}
	\end{equation}
	subject to
	\begin{equation}
		\begin{aligned}
			e_{\left ( 1,1 \right ) }^{XX}  &=\lambda_{1} + \lambda_{3},\\
			e_{\left ( 1,1 \right ) }^{ZZ} &=\lambda_{2} + \lambda_{3},
		\end{aligned}
	\end{equation}

    \begin{equation}
    	\begin{aligned}
    		q_{\text {s}} &=(E_{(\mu, \mu)}^{ZZ})^{b}+(1-E_{(\mu, \mu)}^{ZZ})^{b},\\
    		\tilde{E}_{(\mu, \mu)}^{ZZ} & = \frac{(E_{(\mu, \mu)}^{ZZ})^{b}}{(E_{(\mu, \mu)}^{ZZ})^{b}+(1-E_{(\mu, \mu)}^{ZZ})^{b}},
    	\end{aligned}
    \end{equation}
	\newpage
	and
	\begin{equation}
		\begin{aligned}
			\tilde{\lambda}_{0} & =\frac{(\lambda_{0}+\lambda_{1})^{b}+(\lambda_{0}-\lambda_{1})^{b}}{2 [\left(\lambda_{0}+\lambda_{1}\right)^{b}+\left(\lambda_{2}+\lambda_{3}\right)^{b}]}, \\
			\tilde{\lambda}_{1} & =\frac{(\lambda_{0}+\lambda_{1})^{b}-(\lambda_{0}-\lambda_{1})^{b}}{2 [\left(\lambda_{0}+\lambda_{1}\right)^{b}+\left(\lambda_{2}+\lambda_{3}\right)^{b}]}, \\
			\tilde{\lambda}_{2} & =\frac{(\lambda_{2}+\lambda_{3})^{b}+(\lambda_{2}-\lambda_{3})^{b}}{2 [\left(\lambda_{0}+\lambda_{1}\right)^{b}+\left(\lambda_{2}+\lambda_{3}\right)^{b}]}, \\
			\tilde{\lambda}_{3} & =\frac{(\lambda_{2}+\lambda_{3})^{b}-(\lambda_{2}-\lambda_{3})^{b}}{2 [\left(\lambda_{0}+\lambda_{1}\right)^{b}+\left(\lambda_{2}+\lambda_{3}\right)^{b}] },
		\end{aligned}
	\end{equation}
	where $\tilde{E}_{(\mu, \mu)}^{ZZ}$ represents the overall error rate after the AD method step.

\end{document}